\documentclass[useAMS,usenatbib]{article} \usepackage{graphicx} \input{epsf}
 \usepackage{a4wide}

\def\lesssim{\mathrel{\hbox{\rlap{\hbox{\lower4pt\hbox{$\sim$}}}\hbox{$<$}}}}
\def\gtrsim{\mathrel{\hbox{\rlap{\hbox{\lower4pt\hbox{$\sim$}}}\hbox{$>$}}}}

\begin{document}

\begin{center}

\noindent{\bf \Large Forming Jupiter, Saturn, Uranus and Neptune in Few Million Years by
Core Accretion}

\vspace*{1cm}

Omar G. Benvenuto$^{1,2,3}$, Andrea Fortier$^{1,3,4}$, Adri\'an Brunini$^{1,3,5}$

\end{center}

\vspace*{1cm}

\noindent 1.- Facultad de Ciencias Astron\'omicas y Geof\'\i sicas, Universidad Nacional
de La Plata (UNLP), 1900 La Plata, Argentina.\\

\noindent 2.- Member of the Carrera del Investigador Cient\'\i fico, Comisi\'on de
Investigaciones Cient\'\i ficas de la Provincia de Buenos Aires (CIC), Argentina.\\

\noindent 3.- Instituto de Astrof\'{\i}sica de La Plata (IALP), CCT-Consejo Nacional de
Investigaciones Cient\'\i ficas y T\'ecnicas (CONICET), UNLP,
Argentina.\\

\noindent 4.- Fellow of CONICET.\\

\noindent 5.- Member of the Carrera del Investigador Cient\'\i fico, CONICET.\\

\begin{abstract} Giant planet formation process is still not completely understood. The current most
accepted paradigm,  the core instability model, explains several observed properties of the solar
system's giant planets but, to date, has faced difficulties to account for a formation time shorter
than the observational estimates of protoplanetary disks' lifetimes, especially for the cases of
Uranus and Neptune.  In the context of this model, and considering a recently proposed primordial
solar system orbital structure, we performed numerical calculations of giant planet formation.  Our
results show that if accreted planetesimals follow a size distribution in which most of the mass
lies in 30-100 meter sized bodies, Jupiter, Saturn, Uranus and Neptune may have formed according to the nucleated instability scenario.  The
formation of each planet occurs within the time  constraints and they end up with  core masses in good
agreement with present estimations. \end{abstract}

\noindent   {\it   key   words:}   Planetary   formation; Accretion;    Planetesimals.

\noindent Correspondence should  be directed to: Omar G. Benvenuto, Facultad de Ciencias 
Astron\'omicas y Geof\'\i sicas de La Plata,  Paseo del  Bosque  S/N (1900)  La  Plata, Argentina.  
E-mail: obenvenu@fcaglp.fcaglp.unlp.edu.ar\vskip 0.5 cm 

\newpage

Terrestrial planets are widely believed to form in the inner  solar nebula through the accretion  of
a  population  of centimeter  to kilometer sized planetesimals orbiting the early Sun (Wetherill
1990). The core-accretion model states   that  this  process   also  occurred   in  the   outer  solar
nebula (Mizuno 1980; Pollack et al. 1996).  When a  solid  embryo  has grown  to  few  thousandths 
of  the Earth  mass,  a  gaseous envelope  in hydrostatic  equilibrium begins  to bind  to it.  This 
equilibrium is achieved by  the balance between the heating due to the energy released by  the
incoming planetesimals and the gravity of the core. As  the  core grows  and  planetesimals  exhaust,
the  gaseous envelope eventually can  no longer be supported  and contracts. This  leads to  the last
stage of giant planet formation, a short period  during which the  protoplanet quickly gains the  bulk
of its total mass  (Pollack et al. 1996).   On the other hand, the first stage of the process is
dominated by the formation of the core. The time scale of core growth is strongly dependent on
several  factors, the initial disk surface density of solids in the giant planet region, $\Sigma$, and
the radius of accreted planetesimals being among the most important ones. 

The first calculations that self-consistently considered both the accretion of gas  as well as
planetesimals showed  that, even prescribing  an unrealistically fast core accretion rate, with a
surface density of solids of $\Sigma= 10 \, \mathrm{g} \, \mathrm{cm^{-2}}$ at 5.2~AU, Jupiter takes
6~million years (My) to  complete its  formation (Pollack et al. 1996, updated by Hubickyj et al.
2005)\footnote{This result corresponds to the simulation $10\mathrm{H}^{\infty}$ of Hubickyj et al.
(2005) computed with full interstellar opacities.}, and Saturn  more than 10~My (Pollack et al. 1996).
However, according to observations, the lifetime of circumstellar disks at the Jupiter-Saturn region
is between 3-8 My, while beyond 10~AU they are expected to last somewhat longer (Hillenbrand 2005).
This fact limits the formation time of gaseous planets. Increasing the  local density speeds up  the
formation process, but at  the cost of building too massive cores: interior models of Jupiter that 
agree  with  all observational  constraints show  that  its core mass is at  most
12~$\mathrm{M_\oplus}$ (Guillot 2005), whereas more recent calculations give core masses of
14~-~18~$\mathrm{M_\oplus}$ (Militzer et al. 2008). In the case of Saturn, its core mass is between  
9-22~$\mathrm{M_\oplus}$ (Guillot 2005). Recent  improved  core accretion calculations adopt more
adequate solid  accretion rates for  the cores (Fortier et  al. 2007, 2009),  or  take into  account
local   density   patterns   in  the   disk (Klahr \& Bodenheimer 2006),   but nevertheless they have 
not solved the time scale  problem. Other recent models showed that the time scale for envelope 
growth depends on  the assumed  dust grain  opacity (Ikoma et al. 2000). Grains  entering the
protoplanetary  envelope may coagulate and settle  down quickly.  As a consequence, the correct
opacity  may be far lower  than usually believed (Podolak 2003), easing an early envelope contraction 
that allows the formation of Jupiter in less than 3 My (Hubickyj et al. 2005). However, the  time
scale to form  the  core, one of the  main problems  of  the core accretion model, is  not
substantially modified and, to  date, no self-consistent calculations of the opacity in the
protoplanet's atmosphere have been performed (Movshovitz \& Podolak 2008). Improved solid surface
density calculations (Dodson~-~Robinson et al. 2008) allow to compute formation models of Saturn
consistent with circumstellar disk lifetimes. However, their simulations end with a total mass of
heavy elements higher than current estimations for Saturn.  In all core accretion calculations 
performed so far, a single sized planetesimal    population   has been assumed. However, remarkably,
the accretion rate of solids is strongly dependent  on   the  planetesimal  size.  Therefore, 
self-consistent numerical simulations, in which the core is formed through accretion of
planetesimals following a  size distribution are in order.  At early stages, larger planetesimals
grow faster than smaller ones, resulting in a runaway growth of the    largest (Greenberg et al. 1978;
Wetherill \& Stewart 1989; Kokubo \& Ida 1996).  Runaway bodies then become protoplanetary embryos
that grow by the accretion of small planetesimals. When the mass  of a  protoplanet exceeds a  critical value far lower than the Earth mass,  runaway growth slows  down (Ida \& Makino 1993) and switches to oligarchic growth (Kokubo \&
Ida 1998), in which similar-sized protoplanets dominate the planetesimal system. During the
oligarchic regime only protoplanets grow as there is no substantial accretion among planetesimals. Remarkably, solid bodies in the swarm are not of uniform size but span over a wide interval ranging from millimeter to kilometer.

Numerical simulations show that the mass
distribution of these residual planetesimals may be represented by a single or piecewise power law  $dn_c(m)/dm \propto n(m) \propto
m^{-\alpha}$, where $n_c(m)$ is the number  of bodies larger than a given mass $m$ and $n(m)$ is the
number of bodies in a linear mass bin. For a constant value of $\alpha$  the mass in the interval is $\int n\;  m\; dm \propto
m^{2-\alpha}$. Depending on the value of $\alpha$ the mass in the interval would be mostly contained in small ($\alpha > 2$) or  large ($\alpha < 2$) planetesimals.  
Wetherill and Stewart (1993) studied the evolution of the planetesimal system considering an initial  population of planetesimals whose  radius
is $\sim$~10 km that evolved only through collisions and fragmentation.
They found that planetesimal size distribution relaxes to a piecewise power law: a population of small
planetesimals due to fragmentation ($\alpha \sim 1.7$) and a population of large
planetesimals that follow an accretive regime ($\alpha \sim 2.5$) which, in turn, contained most of the mass of the system. 
Kokubo and Ida (2000) studied through N-body simulations the evolution of planetesimal size and they
found that, in the oligarchic regime, large planetesimals follow a continuous power law distribution
with $2<\alpha < 3$. However, these studies do not take into account the effects due to magneto-rotational instability
turbulence, which establish the  predominance of an erosive rather than an accretive regime for
planetesimal growth, nor the existence of dead zones that could favor accretion
among planetesimals (Ida et al. 2008). On the other hand, recent laboratory experiments show that reaccumulation of fragmentation debris
can lead to the formation of planetesimals (Teiser and Wurm, 2009). These effects could have a substantial impact on the accretion process of planetesimals, specially in the small planetesimal tail of the distribution. 
Evidently, the primordial planetesimal size
distribution in the protoplanetary disk is still an open problem.  In our study, for the sake of
simplicity  we shall consider  a continuous power law distribution, adopting for the exponent the value $\alpha= 2.5$. In view of the present status of the theory of planetesimal growth, we consider
plausible such a distribution.

Another important point to take into account is that the primordial configuration of the outer solar
system was probably  not the one we know today. Numerical simulations (Tsiganis et al. 2005) 
assuming   an  originally  compact planetary system (the Nice model) were able to reproduce the main
observational constraints  regarding the present  orbital structure of the outer solar  system, that
was achieved after  the migration of the fully formed  giant planets to  their present location.   In
$\sim$ 50\% of these simulations Uranus and Neptune switch places, explaining naturally why Neptune's
mass is larger than that of Uranus. The Nice model implies  a solid surface density  profile $\Sigma
\propto a^{-2.168}$,  where $a$ is the distance from the Sun  (Desch 2007).

In our model, in order to be consistent with the Nice model, we placed Jupiter, Saturn, Uranus and
Neptune's embryos at $a= 5.5, 8.3, 14$, and $11$~AU  respectively. We calculate the {\it in situ}
formation of these planets. This assumption, although a simplification of the model, is compatible
with recent studies of the formation of planetary systems. Thommes et al. (2008) found that
solar system analogs come out if the gas giants do not undergo significant migration and remain in
nearly circular orbits. In all our simulations, the initial core and gaseous envelope masses were set
to $\sim 10^{-2} \, \mathrm{ M_\oplus}$ and  $\sim 10^{-10} \, \mathrm{M_\oplus}$ respectively. The
equations  governing  the evolution  of the  gaseous envelope were solved coupled self-consistently 
to the planetesimal accretion rate, employing a standard  finite   difference   (Henyey) method and 
detailed constitutive physics as described in our previous work (Fortier et  al. 2007, 2009). We
adopted the equation of state of Saumon et al. (1995), the grain opacities of Pollack  et al. (1985)
and, for temperatures above $10^3 \; \mathrm{K}$, the molecular opacities of  Alexander \& Ferguson
(1994).

The accretion of planetesimals modify the solid surface density around the planet which, together with
their relative velocities, control the solids accretion rate. The relative velocities are calculated assuming the equilibrium between the gravitational perturbations due to the embryo and the dissipation due to gas drag. 
Since the oligarchic regime dominates
the time scale for the formation of the core, we prescribed this accretion rate during the whole
simulation (Fortier et  al. 2007, 2009). To approximately mimic the effects due to neighboring planets
(Hubickyj et al. 2005), after the planet reaches the first maximum in the solid accretion rate, we
inhibited a further growth of the local planetesimal surface density, i.e., we no longer allow
planetesimals to refill the planetary feeding zone. We assumed that all captured planetesimals sink
onto the solid core and that their mass density is equal to $1.5\; \mathrm{g} \; \mathrm{cm^{-3}}$. We
approached the continuous planetesimal size distribution with a discrete one of nine  sizes  from 
30~m  to  100~km of  radius,  geometrically  evenly spaced. Considering a non homogeneous population
is very important as planetesimal dynamics is affected by nebular drag. Small planetesimals suffer a
stronger damping of random velocities, increasing this way the accretion rate (Safronov 1969); but
they are also subject to faster orbital decay, reducing the planetesimal surface density in the
neighborhood of the protoplanet.  In the nebular model we are using, the accretion time scale of
planetesimals whose radius is $r \geq 30$~m is shorter than the corresponding time scale of orbital
decay (Chambers 2006; Brunini \& Benvenuto 2008). This situation reverts at $r \sim 20$~m.  Therefore,
the giant planets of the solar system should have formed essentially through accretion of
planetesimals larger than $r\sim 30$~m, as smaller planetesimals could not have been  efficiently
accreted. It is worth mentioning that a nebular model based on the present planetary masses, like the
minimum mass solar nebula, does not account for the primordial solid mass comprised in the planetesimals that were not
accreted by the planets. Then, the estimations of the nebular  surface density adopted in this article
only take into account the mass contained in planetesimals that have chances to be accreted.

Fig.~1 shows the results for the formation of Jupiter. Allowing  for a distribution of planetesimal
sizes where the most abundant ones are those of $r= 30 - 80$ m, Jupiter  grows   much  faster  
than   considering   a  typical single size population (e.g. $r=10$ km). Note that the model
corresponding  to   $\Sigma= 11\; \mathrm{g} \; \mathrm{cm^{-2}}$ forms timely and achieves a core
mass  compatible with currently accepted estimations. The  key ingredient that  speeds up planetary 
growth is the  dependence of  the cross-section  of the  planet as a function of the  planetesimal
size.  This, in turn, is determined  by the  drag  force suffered  by  planetesimals  after entering
the planetary envelope, which is proportional to the reciprocal of the planetesimal radius (Podolak et
al. 1988). If there was  no gas bound to the planet, the  core would grow only by  inelastic 
collisions with  planetesimals. For  this process, the cross-section is given  by the projected area
of  the planet enhanced by  gravitational focusing.  Gas drag modifies planetesimal  trajectories, 
eventually trapping them in the planet, greatly enlarging the cross-section (Pollack et al. 1996;
Fortier et  al. 2007, 2009).

Let us consider the conditions for the formation of the giant planets beyond Jupiter's orbit.
Regarding Uranus and Neptune, models consistent with available observations predict a gaseous mass
fraction far lower than those of Jupiter and Saturn. The upper limits for the gas content of Uranus
and Neptune are  of 5.0~$\mathrm{M_\oplus}$ and 4.7~$\mathrm{M_\oplus}$ respectively (Podolak et  al.
2000), while the absolute lower limit for both of  them is  of 0.5~$\mathrm{M_\oplus}$ (Guillot 2005).

Fig.~2 shows the results corresponding to the four giant planets of the solar system computed
separately, together with plausible values for  their core masses (Guillot 2005; Podolak et  al.
2000). Here the four giant planets are formed within  the estimated lifetime of protoplanetary disks.
Other simulations (Alibert et al. 2005) were able to form Jupiter and Saturn in less than 2.5~My but
with Saturn undergoing a migration in conflict with the Nice model (Tsiganis et al. 2005) and
neglecting the oligarchic regime for the growth of the core.   On the other hand, Desch (2007)
estimated the growth time for the cores of the four giant planets in the framework of the
Nice model. Although he did not consider the presence of the planets' atmosphere, his results support
the possibility of the formation of all the giant planets of the solar system in less than 10 My,
provided that small planetesimals were the building blocks of the cores. 

Regarding the mass of the core, as can be seen from Fig.~2, Uranus and Saturn's  fall inside  the
error bar while Jupiter is compatible with the top of its error bar. Neptune's model presents a core
smaller than current estimations.  Nevertheless, we should remark that Neptune's observations are by
far the more inaccurate ones (Podolak et  al. 2000).

 If we consider that the size distribution of planetesimals is represented by 7 species where the
minimum radius is $r=100$ m, instead of 9 species starting with $r=30$ m, the formation time of the
four giant planets is lengthened, being the longest one that of Uranus ($\sim 13$ My). Moreover, 
if we consider that most of the planetesimal mass lies in kilometer-sized bodies, the formation 
time of the ice giants results much longer than the estimated lifetime of the protoplanetary disk.

In order to evaluate the effects of the above mentioned inhibition on the solids accretion rate, we
also computed the growth of the four giant planets without such effect. The results are presented in
Fig.~3. In this case, planetary cores are more massive: Jupiter, Saturn, Neptune and Uranus form cores
of 20, 19, 13 and 11~$\mathrm{M_\oplus}$ respectively. The only core mass outside, but near its error
bar is that of Jupiter. However, in view of the disagreement between different research groups on the
actual value of Jupiter's core mass, we judge our result as acceptable. Also, note that imposing no
inhibition makes the whole planetary formation process even somewhat faster.

 In this study we performed  calculations of giant planet formation in the framework of the nucleated
instability model, taking into account the oligarchic regime for the growth of the core and a
primordial configuration of the solar system according to the Nice model. Also, we considered a size
distribution for the accreted planetesimals in which most of the mass lies in the smallest planetesimals
of the population. This study shows that the formation of the four giant planets of the solar system 
is compatible with the observational constraints imposed by dust disk lifetimes. We found that to
account for the formation of Jupiter, Saturn, Uranus and Neptune in less than 10 My most of the mass
of accreted planetesimals has to be in planetesimals whose radius is $\lesssim 100$ m. We should
remark that if we assume a power law with $\alpha < 2$,  planetary embryos would
grow on a too long timescale. In this sense, we have chosen particularly favorable conditions for the
occurrence of planetary formation within the currently accepted constraints (timescales and core
masses).
 Hopefully, future investigation on the theory of planetesimal formation will provide further insights on the primordial size distribution of planetesimals in the protoplanetary disk.

\noindent{\bf  Acknowledgments}

The authors thank R.E. Martinez and R.H. Viturro for the technical support. 

\vskip 1.0 cm \noindent{\bf  References}

\noindent Alexander, D.~R., Ferguson, J.~W.\ 1994.\ Low-temperature Rosseland opacities.\
Astrophysical Journal 437, 879-891.  

\noindent Alibert, Y., Mousis,  O., Mordasini, C., Benz, W.\ 2005.\ New jupiter and saturn
formation models  meet observations.\ Astrophysical Journal 626, L57-L60.  

\noindent Brunini, A.,  Benvenuto, O.~G.\ 2008.\ On oligarchic growth of planets in
protoplanetary  disks.\ Icarus 194, 800-810.  

\noindent Chambers, J.\ 2006.\ A  semi-analytic model for oligarchic growth.\ Icarus 180,
496-513. 

\noindent Desch, S.~J.\ 2007.\ Mass  distribution and planet formation in the solar nebula.\
Astrophysical  Journal 671, 878-893. 

\noindent Dodson-Robinson, S.~E., Bodenheimer, P., Laughlin, G., Willacy, K., Turner, 
N.~J., Beichman, C.~A.\ 2008.\ Saturn Forms by Core Accretion in 3.4 Myr. Astrophysical 
Journal Letters, 688, L99 

\noindent Fortier, A., Benvenuto, O.~G., Brunini, A.\ 2007.\ Oligarchic planetesimal
accretion and giant planet formation.\ Astronomy and Astrophysics 473, 311-322.  

\noindent Fortier, A., Benvenuto, O.~G., Brunini, A.\ 2009.\ Oligarchic planetesimal
accretion and giant planet formation II.\ Astronomy and Astrophysics, in press (DOI:
10.1051/0004-6361/200811367).  

\noindent Greenberg, R.,  Hartmann, W.~K., Chapman, C.~R., Wacker, J.~F.\ 1978.\
Planetesimals to  planets - Numerical simulation of collisional evolution.\ Icarus 35, 1-26. 

\noindent Guillot, T.\ 2005.\ The  interiors of giant planets: Models and outstanding
questions.\ Annual  Review of Earth and Planetary Sciences 33, 493-530.  

\noindent Hillenbrand, L.~A.\ 2005.\  Observational constraints on dust disk lifetimes:
implications for planet  formation.\ ArXiv Astrophysics e-prints arXiv:astro-ph/0511083.  

\noindent Hubickyj, O.,  Bodenheimer, P., Lissauer, J.~J.\ 2005.\ Accretion of the gaseous
envelope  of Jupiter around a 5~-~10 Earth-mass core.\ Icarus 179, 415-431.  

\noindent Ida, S., Makino, J.\  1993.\ Scattering of planetesimals by a protoplanet -
Slowing down of  runaway growth.\ Icarus 106, 210-227.  

\noindent Ida, S., Guillot, T.,  Morbidelli, A.\ 2008.\ Accretion and Destruction of
Planetesimals in  Turbulent Disks.\ Astrophysical Journal 686, 1292-1301.  

\noindent Ikoma, M., Nakazawa, K.,  Emori, H.\ 2000.\ Formation of giant planets:
dependences on core accretion  rate and grain opacity.\ Astrophysical Journal 537,
1013-1025.  

\noindent Klahr, H.,  Bodenheimer, P.\ 2006.\ Formation of giant planets by concurrent
accretion  of solids and gas inside an anticyclonic vortex.\ Astrophysical Journal  639,
432-440. 

\noindent Kokubo, E., Ida, S.\  1996.\ On runaway growth of planetesimals.\ Icarus 123,
180-191.  

\noindent Kokubo, E., Ida, S.\  1998.\ Oligarchic growth of protoplanets.\ Icarus 131,
171-178. 

\noindent Kokubo, E., Ida, S.\  2000.\ Formation of protoplanets from planetesimals in the
solar nebula.\  Icarus 143, 15-27. 

\noindent Militzer, B., Hubbard,  W.~B., Vorberger, J., Tamblyn, I., Bonev, S.~A.\ 2008,
Astrophysical  Journal Letters, 688, L45  

\noindent Mizuno, H.\ 1980.\ Formation of  the giant planets.\ Progress of Theoretical
Physics 64, 544-557. 

\noindent Movshovitz, N.,  Podolak, M.\ 2008.\ The opacity of grains in protoplanetary
atmospheres.\  Icarus 194, 368-378. 

\noindent Podolak, M.\ 2003.\ The contribution of small grains to the opacity of
protoplanetary atmospheres.\  Icarus 165, 428-437. 

\noindent Podolak, M., Pollack,  J.~B., Reynolds, R.~T.\ 1988.\ Interactions of
planetesimals with  protoplanetary atmospheres.\ Icarus 73, 163-179. 

\noindent Podolak, M., Podolak, J.~I., Marley, M.~S.\ 2000.\ Further investigations of
random models of Uranus and Neptune.\ Planetary and Space Science 48, 143-151.  

\noindent Pollack, J.~B., McKay, C.~P., Christofferson, B.~M.\ 1985.\ A calculation of the
Rosseland mean  opacity of dust grains in primordial solar system nebulae.\ Icarus 64,
471-492.  

\noindent Pollack, J.~B.,  Hubickyj, O., Bodenheimer, P., Lissauer, J.~J., Podolak, M.,
Greenzweig,  Y.\ 1996.\ Formation of the giant planets by concurrent accretion of solids 
and gas.\ Icarus 124, 62-85. 

\noindent Safronov, V.S. \ 1969. \ Evolution of of the  protoplanetary cloud and formation
of the earth and planets. Nauka, Moscow. 

\noindent Saumon, D., Chabrier, G., van Horn, H.~M.\ 1995.\ An Equation of State for
Low-Mass Stars and  Giant Planets.\ Astrophysical Journal Supplement Series 99, 713.  

\noindent  Teiser, J., Wurm, G.\ 
2009.\ High-velocity dust collisions: forming planetesimals in a 
fragmentation cascade with final accretion.\ Monthly Notices of the Royal 
Astronomical Society 393, 1584-1594.  

\noindent Thommes, E.~W.,  Matsumura, S., Rasio, F.~A.\ 2008.\ Gas disks to gas giants:
Simulating the  birth of planetary systems.\ Science 321, 814-817.  

\noindent Tsiganis, K., Gomes,  R., Morbidelli, A., Levison, H.~F.\ 2005.\ Origin of the
orbital  architecture of the giant planets of the Solar System.\ Nature 435,  459-461.  

\noindent Wetherill, G.~W.\ 1990.\  Formation of the earth.\ Annual Review of Earth and
Planetary Sciences 18,  205-256.

\noindent Wetherill,  G.~W., Stewart, G.~R.\ 1989.\ Accumulation of a swarm of small 
planetesimals.\ Icarus 77, 330-357.  

\noindent Wetherill,  G.~W., Stewart, G.~R.\ 1993.\ Formation of planetary embryos - Effects of fragmentation, low relative velocity, and independent
variation of eccentricity and inclination \ Icarus 106, 190-209.  


\newpage

\begin{figure} \centering \includegraphics[angle=0, width=0.7\textwidth]{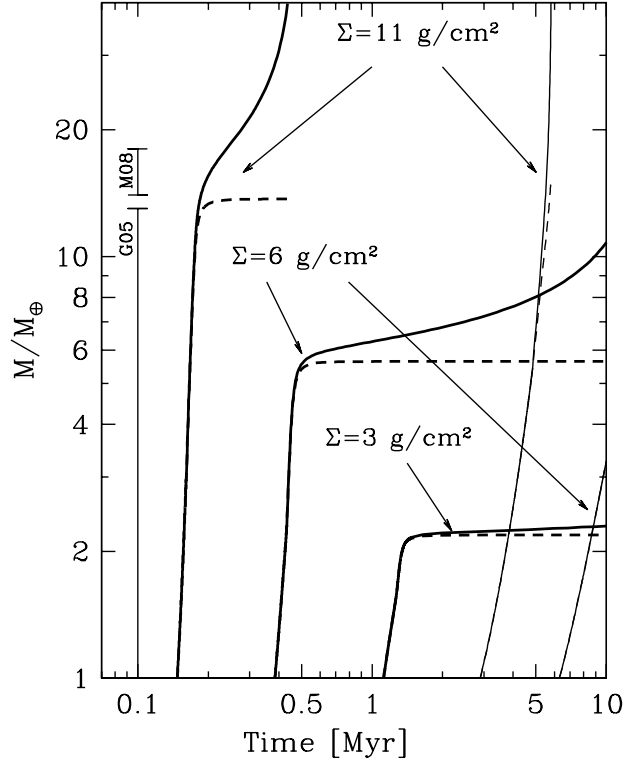} 
\caption{ The growth of Jupiter for different  surface densities.  Thick solid
(dashed) lines depict the growth of the total (core) mass for models computed with a size
distribution for the accreted planetesimals. Thin lines stand for the case of a single sized
population  of planetesimals with $r= 10$~km.  The error bars depict the currently accepted
values for Jupiter's core mass. G05 denotes the values given by Guillot (2005) while M08 stands
for the results presented by Militzer et al. (2008). $\Sigma$ represents the assumed values for
the initial disk surface density of solids in the Jupiter region.  Note that the formation is a
two step process. First, the core grows by planetesimal accretion, whose gravitational energy
release inhibits the gaseous envelope from undergoing a premature contraction. When the
neighborhood of the protoplanet is depleted from planetesimals, the core mass gets its
asymptotic value and a runaway gas accretion is established. Quickly, the planet reaches its
final mass (318~$\mathrm{M_\oplus}$). Clearly, the considered size distribution of
planetesimals yields a formation time scale much shorter than considering a single planetesimal
size of $r= 10$~km. In particular, with a local surface density of $11 \; \mathrm{g}
\;\mathrm{cm^{-2}}$, we obtain an acceptable  core mass value and a comfortably short formation
time. Therefore we assume that this result scales the density throughout the entire protoplanetary
disk.} 
\end{figure}

\begin{figure} \centering \includegraphics[angle=0, width=0.8\textwidth]{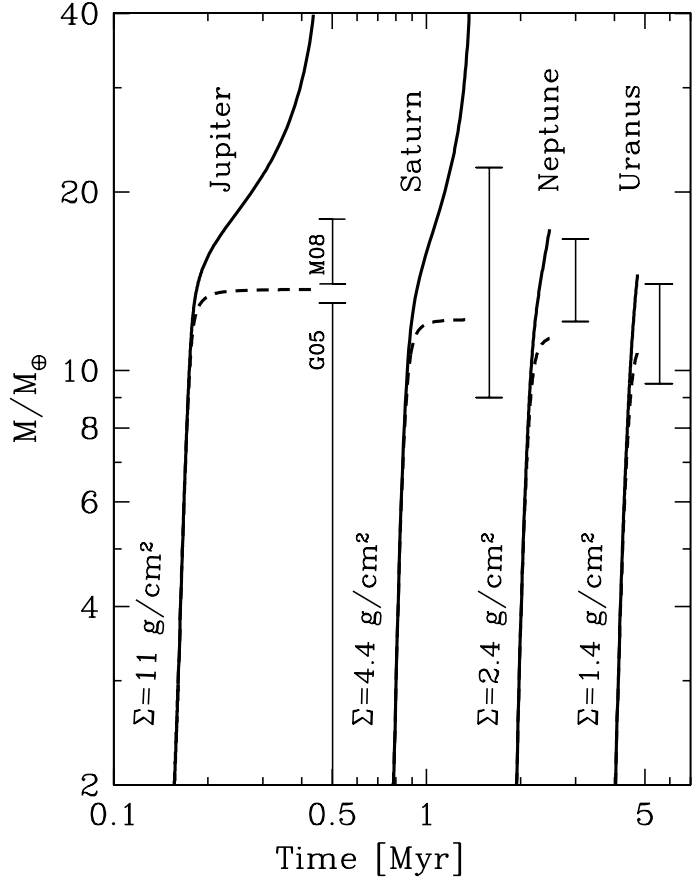} 
\caption{ The growth of the four giant planets. Solid (dashed) lines depict the
growth of the total (core) mass for models computed with a size distribution for the accreted
planetesimal. The values of $\Sigma$ correspond to the initial surface density of solids at the
position of each embryo. The error bars depict the currently accepted values for the core
masses of each planet. For the case of Jupiter, as in Fig. 1, the lower error bar is that of
Guillot (2005), whereas the upper one corresponds to the results given by Militzer et al.
(2008). Simulations were stopped when the final mass of each planet was reached (approximately
318, 95, 14, and 17~$\mathrm{M_\oplus}$ for Jupiter, Saturn, Uranus and Neptune respectively).
The formation times are: Jupiter 0.44~My, Saturn 1.4~My, Neptune 2.5~My and Uranus 4.75~My.} 
\end{figure}

\begin{figure} \centering \includegraphics[angle=0, width=0.8\textwidth]{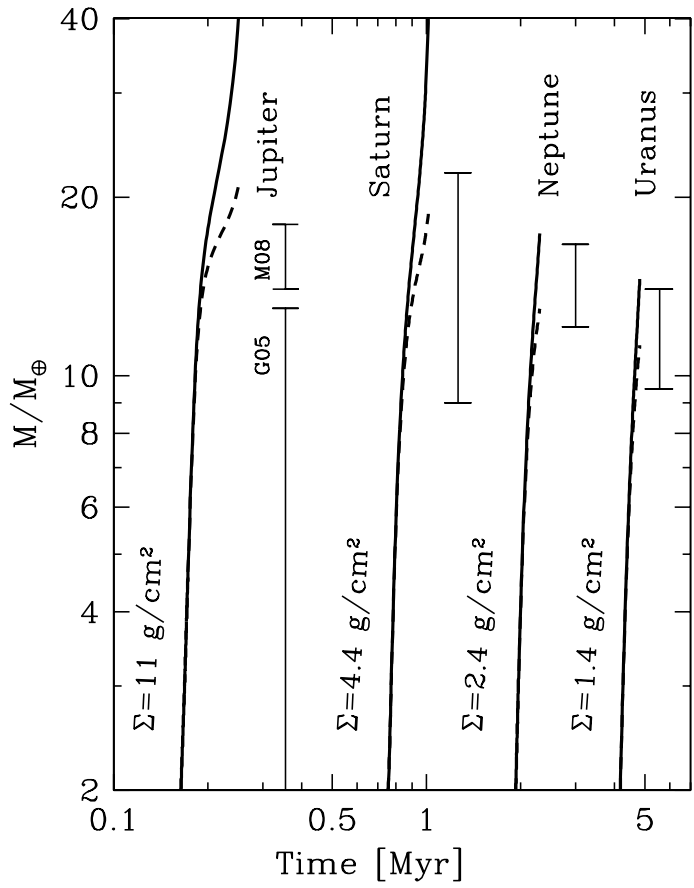} 
\caption{ Same as  Fig.~2 but without imposing any inhibition on the solid mass accretion rate.}
\end{figure}

\end{document}